\documentclass[11pt,leqno,twoside]{amsart}

\usepackage{enumerate}

\usepackage{amssymb}

\usepackage[english]{babel}
\usepackage{amssymb,amsthm,amsmath,eucal,mathrsfs}
\usepackage{bm}

\usepackage{amsmath}
\usepackage{amsfonts}
\usepackage{amsmath}
\usepackage{amssymb}
\usepackage{graphicx}
\usepackage{lscape}
\usepackage{amstext}
\usepackage{amsthm}
\usepackage{color}
\usepackage{float}
\usepackage{mathrsfs}
\usepackage{epsfig}
\usepackage{url}
\usepackage{fancyhdr}
\usepackage{pspicture}
\usepackage{graphicx}

\usepackage{cite}

\usepackage{amsmath,verbatim}

\usepackage{amsthm}

\usepackage{amssymb}

\usepackage{amsfonts}

\usepackage{dsfont}

\usepackage{hyperref}

\newtheorem{theorem}{Theorem}[section]

\theoremstyle{definition}

\newtheorem{definition}[theorem]{Definition}

\newtheorem{remark}[theorem]{Remark}

\numberwithin{equation}{section}

\renewcommand{\Im}{{\ensuremath{\mathrm{Im\,}}}} 

\renewcommand{\Re}{{\ensuremath{\mathrm{Re\,}}}} 

\renewcommand{\div}{\mathrm{div}\,}    

\newcommand\restr[2]{{
  \left.\kern-\nulldelimiterspace 
  #1 
  \vphantom{\big|} 
  \right|_{#2} 
  }}

\usepackage{eucal}

\title[Mathematics Of The Imaging Modalities Using Small Scaled Contrast Agents]{An Introduction To The Mathematics Of The Imaging Modalities Using Small Scaled Contrast Agents}

\author[Ghandriche and Sini]{ Ahcene Ghandriche$^*$ and Mourad Sini$^{\ddag}$
}
\thanks{$^*$ RICAM, Austrian Academy of Sciences, Altenbergerstrasse 69, A-4040, Linz, Austria. Email: ahcene.ghandriche@ricam.oeaw.ac.at. This author is supported by the Austrian Science Fund (FWF): P 30756-NBL}
\thanks{$^{\ddag}$ RICAM, Austrian Academy of Sciences, Altenbergerstrasse 69, A-4040, Linz, Austria. Email: mourad.sini@oeaw.ac.at. This author is partially supported by the Austrian Science Fund (FWF): P 30756-NBL}

\begin{document}

\date{}

\allowdisplaybreaks

\begin{abstract} 
 
	In the recent years, we witness a great interest in imaging, in a wide sense, using contrast agents. 
	One of the reasons is that many imaging modalities, as the ones related to medical sciences, suffer from several shortcomings. The most serious one is the issue of instability. Indeed, it is, nowadays, a common certainty that  classical inverse problems of recovering objects from remote measurements are, mostly, highly unstable. To recover the stability, it is advised to create, whenever possible, the missing contrasts in the targets to image. In this survey paper, we follow this direction and propose an approach how to analyze mathematically the effect of the injected agents on the different fields under consideration. These contrast agents are small-sized particles modeled with materials that enjoy high contrasts as compared to the ones of the  background. These two properties allow them, under critical scales of size/contrast, to create local spots when excited from far. These local spots can be remotely recovered in stable ways. The accessible information on the target are encoded in theses spots.     
	After stating a class of such imaging modalities that enter into this framework, as the acoustic imaging, photo-acoustic imaging, optical imaging and more, we provide detailed analysis for first two modalities where the contrast agents are micro-bubbles and nano-particles respectively. In these cases, we provide a clear and useful correspondence between the critical size/contrast scales and the main resonances, and hence the local spots, they are able to create while excited with appropriate incident frequencies. To estimate the remote dominant field generated by the background in the presence of such particles, we derive the point-interaction approximation of these fields. This dominant field that we call the Foldy-Lax field, as it is reminiscent to the Foldy-Lax field generated by Dirac-like potentials with prescribed multiplicative (scattering) coefficients, encodes the fields after the multiple scattering between the background and the different particles. This Foldy-Lax field contains the accessible information on the target to image. Using resonating incident frequencies enhances this field and makes it readable from remote measurements.           
\end{abstract}

\subjclass[2010]{35R30, 35C20}
\keywords{Mathematical imaging, contrast agents, micro-bubbles, nano-particles, Neumann-Poincar\'e operator, Newtonian operator, Minnaert resonance, plasmonic resonances, dielectric resonances.}

\maketitle


\setcounter{tocdepth}{1}
  \tableofcontents

\section{Introduction and motivation} \label{Introduction}

Our major focus in this survey is the mathematical analysis of problems described by partial differential models involving micro or nano-scaled structures enjoying high contrasts as compared to their surrounding background.
The motivations of these studies come from inverse problems, medical imaging and material sciences.  Here, we are mainly concerned with the inverse problems related to medical imaging. 
 In a large class of inverse and imaging problems that have been studied in the recent decades, the goal is to use remotely measured data
 to reconstruct internal features of the mechanism that produced them. In general, the mechanism is described by partial differential equations with material coefficients entering them.
 Estimating (few of) these coefficients is the target of these problems. What we learned in the last two decades is that such problems are highly unstable, meaning that a small 
 perturbation of the measured data can produce a large deviation in the reconstructed coefficients. These instabilities are unavoidable as they are inherent to the settings themselves.
 The main reason is that the forward maps are usually highly smoothing. There are several ways that one can follow to handle the issue of instability. 
 The first approach is to try to stabilize such inverse and imaging problems. One way to stabilize them is to approximate them by a sequences of stable ones and see how the solutions of the stable ones 
 converge to the true one. This is called the regularization approach, see \cite{E-H-N-2000} for more details. 
  This approach of stabilizing the problems is a purely mathematical idea created since A. N. Tikhonov and J. L. Lions and passing by A. Calderon, see \cite{Isakov-book} for an overview on different ideas proposed for solving these ill-posed problems. 
 The other approach, which is purely an engineering idea, is to go back to the data (the measured signals) and enhance them. One way of enhancing the signals is to inject small-scaled contrast agents into the target to image. 
 Our goal is to analyze mathematically this approach, but most importantly to find how to quantify it and use it to solve the related inverse and imaging problems. We focus on imaging modalities involving 
 micro-bubbles (as in Acoustics) and nano-particles (in models related to the Electromagnetism). There are of course other highly important and challenging models as the Elasticity and the models derived as combination of these models in the framework of multiphysics imaging (as Photo-Acoustic imaging, Magneto-Acoustic imaging, MREIT, etc), see \cite{Habib-book} for further discussions.
\bigskip

Borrowing similar comments from the related experimental engineering literature, as a support to the second approach, it is well known that conventional imaging techniques, as microwave imaging techniques, are potentially capable of extracting features in breast cancer, for instance, in case of the relatively high contrast 
between healthy tissues and malignant ones, \cite{F-M-S:2003}. However, it is observed that in case of benign tissue, the variation of the contrasts is quite low so that such conventional imaging modalities are limited to be used for early detection of such diseases. Creating such missing contrasts is highly desirable. One way to do it is to use either electromagnetic nano-particles, see \cite{B-B:2011}, or micro-bubbles, see \cite{Q-C-F:2009}, 
as contrast agents. 

Next, we describe a list of few imaging modalities that could fit into the strategy we are willing to follow.

\begin{enumerate}
 \item Acoustic Imaging using micro-bubbles as contrast agents, see \cite{Abderson-al-2011, Q-C-F:2009, Quaia-2007, Ilov-al-2018} for more details on related theoretical and experimental studies. This modality is based on using the contrasted scattered fields, by the targeted anomaly, measured before and after injecting micro-scaled bubbles. These bubbles are modeled by mass densities and bulk moduli enjoying contrasting scales. These contrasting scales allow them to resonate at certain incident frequencies. The goal then is to analyze mathematically this contrasted scattered fields in terms of these scales with incident frequencies close to these resonances and derive explicit formulas linking the values of the unknown mass density and bulk modulus of, the targeted region, to the measured scattered fields.    
 
\bigskip
 
 \item Optical Imaging using electromagnetic nano-particles as contrast agent, see \cite{B-B:2011, Challa-Choudhury-Sini, F-M-S:2003}. In the same spirit as for the Acoustic imaging, we contrast the measured electromagnetic field before and after injecting the nano-particles. In the original works, see \cite{B-B:2011} for instance, the authors propose to inject a 'relatively dense' set of magnetic (or dielectric) nano-particles. Using effective medium theory, we can transform the imaging problem into a particular inverse scattering problem. In addition, instead of using a dense set of such nano-particles, we propose to inject few of them but we choose them to be nearly resonating nano-particles. Precisely, in this case, we need to use frequencies of incidence close to the corresponding \emph{plasmonic or dielectric resonances (i.e. Mie resonances)}. Hence the equivalent scattering problem enjoys local coercivity which enables convexity of the related energy functional. 
 \bigskip
 
 \item Photo-Acoustic Imaging using magnetic or dielectric nano-particles as contrast agents, see \cite{L-C:2015, P-P-B:2015}. Exciting injected magnetic (or dielectric) nano-particles in the targeted tissue will create heat on their surfaces (or bodies) and in its turn the heat will create pressure waves. The imaging problem is then to recover the permittivity from these remotely measured pressure waves. 
 We have two major steps for this imaging modality. The first concerns the photo-acoustic inversion. In this step, we recover the electric energy created by the nano-particles. 
 The  second step is the electromagnetic inversion: from the reconstructed electric energy, we recover the permittivity (and eventually the conductivity). Here also, \emph{the plasmonic or dielectric resonances} will play a key role.

 \bigskip
 
 \item Magnetic Particle Imaging. Created in early 2000, see \cite{Gleich}, 
 the method is based on tracking injected volume of magnetic nano-particles diluted in a fluid. 
 Indeed, it is quantitative imaging method which uses the nonlinear remagnetization of the used nano-particles to determine their location, see \cite{MPI-book} for more details.
 Hence the resulting MRI images, i.e. the measured potentials, are used to evaluate the density of distribution of the nano-particles. The reason why the discrepancy between the scattered field and the incident field becomes high, and hence useful, is due to the nonlinear character of the remagnetization (which is due to the nonlinear susceptibilities of the used nano-particles). We would like to study mathematically this imaging modality to
 \begin{enumerate}
 
\item quantify the link between the measured potentials and the density of distribution of the nano-particles.  
 
\item instead of using the nonlinear character to enhance the scattered fields, we propose to use nearly resonant particles.
The advantage is that we do not need to use a high concentration of the nano-particles to be able to track them. 
This will reduce the amount of particles to be injected into the targeted body.
 \end{enumerate}
 There are few mathematical works devoted to this method, as \cite{MPI-1} and \cite{MPI-2} for instance.
 \bigskip
 
 \item Nuclear Imaging: Hadron-therapy. Contrary to the previously described imaging modalities, the target of this method is to be a final therapy for patients at their final stages 
 (i.e. for extremely advanced tumors). The main idea is to accelerate the protons at very high speed to reach very high energies before to deliver them to the part of the body to be cured. 
 The advantage, as compared to X-rays based methods, is that the energy is spread only on or close to the tumor while the X-rays dissipate energy while traveling. The goal is to understand how it works by localizing and quantifying the amount of energy transported near the tumor. Here also, to our best knowledge, there is no 
 mathematical work devoted to analyzing this method, see \cite{Hadron-therapy-1, Hadron-therapy-2} for more detailed information.

\end{enumerate}
\bigskip

\begin{figure}[h]
\includegraphics[scale=0.5]{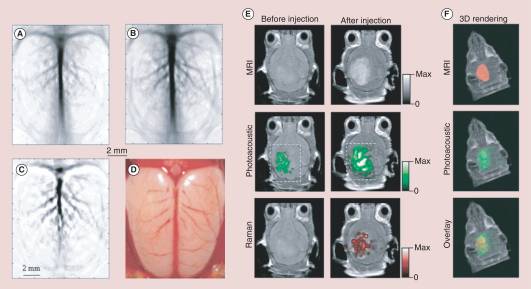}
\caption{\textit{Non-invasive photo-acoustic imaging of a rat's cerebral cortex using nano-particles as contrast agents. This figure is from \cite{L-C:2015}}.}
\label{f1}
\end{figure}

\begin{figure}[h]
\includegraphics[scale=0.5]{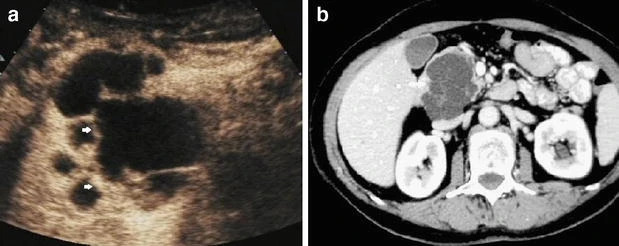}  
\caption{\textit{(a) Multilocular pancreatic cystic mass revealing intracystic septal enhancement $45$ s after micro-bubble injection with ultrasound. (b) The pattern is confirmed at contrast-enhanced CT. This figure is from \cite{Quaia-2007}}.}
\label{f2}
\end{figure}

In the sequel, to describe our approach, we will mainly focus on the acoustic imaging using micro-bubbles, i.e. (1), and the photo-acoustic imaging using (dielectric) nano-particles, i.e. (3). In Figure \ref{f1} and Figure \ref{f2}, we see the sharpness of the images when injecting nano-particles or micro-bubbles respectively in the targeted regions. This suggests, in particular, that contrasting the images taken before and after injecting the small-scaled contrasts agents would allow us to extract quantitative information on the targeted regions, as the mass density and the bulk modulus, for the acoustic imaging, or the electric permittivity, and eventually the conductivity, for the photo-acoustic imaging of the targeted region. 
Our aim is to understand and quantify this.  
\bigskip

At the mathematical analysis level, we need to study how the acoustic pressure and the electromagnetic waves (for the related imaging modalities) are perturbed by the presence 
of micro-bubbles or nano-particles having highly singular relative densities and/or bulks or relative electric permittivity or relative magnetic permeability. 
For both the two imaging modalities we described above, we need \emph{to handle non periodic distributions} of the small scaled inclusions and hence homogenization type techniques do not apply. In addition, of a particular importance to us is the use of micro and nano-scaled inclusions which are \emph{nearly resonating}. In our settings, by resonance we mean the frequencies of incidence for 
which the corresponding forward problem has non-unique solutions. They are solutions of nonlinear dispersion equations. Basically, in such situations, any solution is a linear combination of one which has a finite energy (fixed by 
the radiation condition) and others which are localized near the small inclusions. In general, and surely in our settings, such resonances are located in the (lower half of
the) complex plane. But if we use micro or nano-scaled inclusions which have, in addition, high contrasts, then these resonances are, in fact, close, in terms of the scales and contrasts, to the real line. 
This fact will be extensively used in both the imaging modalities. 

The analysis is based on integral equation methods (both direct and indirect representations), taking into account the different scales related to the size, contrasts and the eventual cluster of the particles. We highlight few of the key arguments in analyzing these imaging modalities:
\bigskip

\begin{enumerate}

\item Rewriting the related Lippmann-Schwinger integral equation for the acoustics (or system for the electromagnetics) as a linear combination of the surface double layer operator (or the Neumann-Poincar\'e operator) and the volume integral operator (or the Newtonian operator), we show that the eventual resonances split into two families: one given by surface-like eigenfunctions and the other body-like eigenfunction corresponding to the eigenvalues of the Neumann-Poincar\'e  and the Newtonian operators respectively. For instance, we have the Minnaert resonance which corresponds to the value $-\frac{1}{2}$ as the eigenvalue of the Neumann-Poincar\'e operator, for acoustics, and the plasmonic resonances related to the eventual sequence of eigenvalues to the same operator for the electromagnetism. On the other hand, the dielectric (or Mie) resonances correspond to the eigenvalues of the Newtonian operator for electromagnetism. These resonances can be, approximately, excited if the material properties of the particles enjoy precise scales of their contrasts with the surrounding background. Such scales will be discussed in section \ref{section2}.  

\item The Neumann-Poincar\'e operator, or the double layer operator, appears due to the contrast of the higher order coefficient as the density for the acoustics model or the permeability for the electromactism model when written solely with the electric fields. It is then not surprising that the dominant terms of the surface-like resonances, i.e. Minnaert for acoustics and plasmonics for electromagntics, depend on that contrast. This is indeed the case as it will shown in section \ref{section4}. This feature will be of importance in acoustic imaging to reconstruct the mass density.
Contrary to these resonances, the volume-like ones (as the dielectric resonances) have their dominant parts independent on the corresponding contrasts (i.e. the contrast of the permittivity and eventually the conductivity).  

\item  The distribution of the small particles is not necessarily periodic, as we handle single particles, doubles (dimers, i.e. closely spaced two by two particles) and also clusters of highly dense particles. Therefore, homogenization techniques are not useful at this point. Rather, the method of analysis we use has its roots back to the Foldy formal method, see \cite{Foldy-1945, Martin-book}, of representing the acoustic field, due to multiple point-like potentials. The multiplicative coefficients attached to these point-like potentials model the scattering strengths (that we call the scattering coefficients). The close form of the Foldy field, solution describes the field generated after all the mutual interaction between the point-like scatterers. Such a representation of the fields has been justified by Berezin and Faddeev in the frame work of quantum mechanics, see \cite{B-F-1961}. The idea is that based on
the Krein extension theory of self-adjoint operators, one can model the diffusion by point-like
particles by the Schr\"oedinger model with singular potentials of Dirac type supported on those
point-like scatterers. This opened different and fertile directions of research related to modeling with singular potentials supported on point, see \cite{Alberverio.al-2005}, lines \cite{E-K-2008} or generally hypersurfaces \cite{M-P-S-2016}. Going back to our subject, and as our small-scaled particles enjoy high contrasts, it is natural then to expect that the dominant part of the field generated by their mutual interactions is reminiscent to the Foldy field. Indeed, this dominating field can be seen as a Foldy field generated by the point-like potentials, centered at the centers of our small particles, where the attaching coefficients, i.e. the scattering coefficients in Foldy's language, are described by a combination of the used incident frequencies and the relative contrasts of the material (mass density for acoustics or permeability for electromagnetism for instance). This combination is nothing but the dominant part appearing in the (nonlinear) dispersion equation. These scattering coefficients can be large when the incident frequency is taken close to the different resonances mentioned above (depending of the scales of our model). This enhancement of the scattered field is the key step in all the imaging modalities we have cited above. To give a taste on how this is used, let us mention the following three situations:

\bigskip

\begin{enumerate}
\item Injecting one single particle, the Foldy field reduce to one element, as there is no multiple scattering. 
In acoustic imaging, we can see from the remotely measured acoustic field, i.e. the far-field for instance, that its value changes drastically whether or not the used incident frequency is close to the resonance (the Minnaert resonance in this case). This discrimination allows us to estimate this resonance. From this resonance, we can estimate the mass density of the background in the vicinity of the injected bubble, see \cite{D-G-S-2020}. In photo-acoustics, we could estimate the internal modulus of the total field (i.e. phaseless total field). This allows us to transform the photo-acoustic imaging to the inversion of the phaseless total electric field, see \cite{AhceneMourad}.   

\item Injecting dimers, i.e. two close particles, then the Foldy field encodes the multiple scattering field between the two particles. From these fields, we can extract not only the total fields on the two centers, but also the background Green's function on the two centers. This Green's function evaluated on the two centers encodes the values of the lower coefficients of our models, i.e. the bulk modulus for acoustic imaging and the permittivity (and eventually the conductivity) for the photo-acoustic imaging, see \cite{AhceneMourad}. However, extracting such information is possible only if the used incident frequencies are close to the mentioned resonances. This issue is related to the Foldy-Lax paradigm, an outstanding open question, on the possibility of extracting, from remote data (i.e. far-fields), the fields generated after multiple scattering between the particles. This paradigm makes sense and it is justified if the used incident frequencies are close to the resonances, see \cite{G-S-2021}. 

\item Injecting a dense cluster of small particles, the corresponding Foldy field approximates the Lippmann-Schwinger field (i.e. the field solution of the Lippmann-Schwinger equation) with a (contrasting) coefficient given by the average of the mutual scattering coefficients of each particle. The sign of this contrast is fixed by the used incident frequency if it is chosen near to the mentioned resonances. As in the metamaterial theory, up to an appropriate choice of the incident frequency, we generate negative lower order coefficients of the effective medium (bulk modulus for acoustics and permittivity for electromagnetism for instance). Hence the derived effective medium enjoys local coercivity. This local coercivity makes the inverse scattering problem more stable. In particular, the related least-square functional might be convex. Again, recovering such nice properties is possible only if the incident frequencies are close to the mentioned resonances.    
\end{enumerate}    

\end{enumerate}


\bigskip

The rest of the manuscript is divide as follows. In section \ref{section2}, respectively section \ref{section3}, we give a more precise description of the kind of micro-scaled, respectively nano-scaled, inclusions we will be using and describe qualitatively and quantitatively the generated resonances at the precise needed scales. The formal characterization of the resonances described in this section is well justified in deriving the needed asymptotic expansions of the different related fields. This is stated in section \ref{section4} for the acoustic imaging and in section \ref{section5} for the photo-acoustic imaging modalities respectively. From these expansions, we describe imaging functional that can be used to effectively extract the needed coefficients from those expansions.


\section{Characterization of the contrast agents}\label{section2}

\subsection{Electromagnetic Nano-particles} 
Different types of nano-particles are proposed in the literature. Let us cite few of them: 

\begin{enumerate}
\item To create contrast in the permittivity, carbon nano-tubes, ferroelectric nano-particles and the calcium copper 
titanate are used, see \cite{B-B:2011}. Such particles have diameter which is estimated around $10$ nm, or $10^{-8}$ m, and have the relative electrical permittivity of the order $10$ for the carbon nanotubes, $10^3$ for ferroelectric nano-particles and around $10^6$ for the Calcium copper titanate, see \cite{W1}. If the benign tumor is located at the cell level (which means that our imaging target $\Omega$ is that cell), with diameter of order $10^{-5}$m, then the $\Omega$-relative radius of the particles is of the order $a \sim 10^{-3}$. Hence, the relative permittivity of the types of nano-particles are estimated of the order $a^{-r}$ where $r$ is $\frac{1}{3}$ for the carbon nanotubes, $1$ for ferroelectric nano-particles and $2$ for the calcium copper titanate.

\item  The human tissue is known to be nonmagnetic. To create magnetic contrasts, it was also proposed in \cite{B-B:2011} to use iron oxid magnetic nano-particles for imaging early tumors.  Such material has a relative 
 permeability of the order between $10^{4}$ and $10^{6}$, see \cite{Hyper-physcs}, and hence of the order $a^{-r}$ with $r$ between $\frac{4}{3}$ and $2$.

\item Other types of material are given through more involved electric or magnetic susceptibilities. Examples of such nano-particles are those for which the permittivity (similarly the permeability) follow spatial dispersion relation as the ones given by the Drude model, i.e. $\epsilon:=\epsilon_0 -\frac{\omega_p^{2}}{\omega(\omega +i \gamma)}$ $\left( \text{or} \; \mu:=\mu_0 -\frac{\omega_p^{2}}{\omega(\omega +i \gamma)} \right)$ with $\omega_p$ as the plasma frequency, $\gamma$ the damping parameter\footnote{Damping parameter is a dimensionless constant.} and $\omega$ is our incident frequency\footnote{The coefficients with zero as subscripts refer to the background media.}. 
\end{enumerate}

This shows that for the detection of the tumors using such nano-particles, we can model
the ratio of the relative electric permittivity and relative magnetic permeability in terms of the 
relative size of nano-particles. Inspired on this, we set the following definition:

\begin{definition}
We call $(D_m, \epsilon_m, \mu_m)$ an electromagnetic nano-particle of shape $ D_m$ with diameter $a$, of order of few tens of nanometers, and permittivity and  permeability $\epsilon_m, \mu_m$ respectively. We call them
\bigskip
\begin{enumerate}
\item \emph{Electric (or Dielectric) Nano-particles} if in addition: $\frac{\epsilon_m}{\epsilon_0}\sim a^{-r}, r>0$ and $\frac{\mu_m}{\mu_0}\sim 1$ as  $a\ll1$.\\
 This implies that the relative index of refraction is large, i.e. $ \frac{\kappa^2_{m}}{\kappa^2_0}:= \frac{\epsilon_{m} \mu_{m}}{\epsilon_{0} \mu_{0}} \gg 1
\mbox{ as } a\ll1$. Hence the relative speed of propagation $\frac{c_m}{c_0}:=\frac{\kappa_0}{\kappa_m}$ is small. But, the contrast of the transmission coefficient is moderate. 
\bigskip


\item  \emph{Magnetic (or Plasmonic) Nano-particles} if in addition $\frac{\epsilon_m}{\epsilon_0} \sim 1$  and $\frac{1}{2}\frac{\mu_m+\mu_0}{\mu_m-\mu_0}$ 
 is ``very close'' to one of the eigenvalues of the Neumann-Poincar\'e operator (i.e. the adjoint of the double layer operator). This means that the relative speed of propagation is moderate. But the contrast of the transmission coefficient is large. Example of such materials are given by the Drude model described above. Indeed, if $\mu_m:=\mu_0 -\frac{\omega^{2}_p}{\omega(\omega +i \gamma)}$, then $\frac{1}{2}\frac{\mu_m+\mu_0}{\mu_m-\mu_0}=\frac{1}{2}-\frac{\mu_0 \, \omega (\omega+i\gamma)}{\omega^{2}_p}$. For smooth shapes $D_m$, the corresponding double layer operator has its spectrum as $\{\sigma_n\}_{n \in \mathbb{N}} \subset [-\frac{1}{2}, \frac{1}{2})$ with $0$ as the only accumulation point for the sequences $(\sigma_n)_{n \in \mathbb{N}}$. Hence the incident frequency $\omega$ and the damping parameter $\gamma$ can be chosen so that $\frac{1}{2}\frac{\mu_m+\mu_0}{\mu_m-\mu_0}$ approaches elements of the sequence $(\sigma_n)_{n \in \mathbb{N}}$.
\end{enumerate}
\end{definition}



\subsection{Micro-Bubbles} 
Similar as for the electromagnetic nano-particles, we set the following definition of the micro-bubbles

\begin{definition}
We call $(D_m, \rho_m, k_m)$ a micro-bubble of shape $D_m$ with diameter $a$,
of about few tens of micrometers, and mass density and bulk modulus $\rho_m, k_m$ respectively. They are called 

\bigskip

\begin{enumerate}
\item \emph{Low Dense / Low Bulk Bubbles} if in addition: $\frac{\rho_m}{\rho_0}\sim a^{r}$ and $\frac{k_m}{k_0}\sim a^{r}$ with $ r>0$ and then  $ \frac{c^2_{m}}{c^2_0}:= \frac{\rho_{m}}{k_{m}} \, \frac{\rho_{0}}{k_{0}}  \sim 1  \mbox{ as } a\ll1$. This means that the relative speed of propagation $\frac{c^2_{m}}{c^2_0}$ is moderate. But the contrast of the transmission coefficient is large.
\bigskip

\item \emph{Moderate Dense / Low Bulk Bubbles} if in addition: $\frac{\rho_m}{\rho_0}\sim 1$ and $\frac{k_{m}}{k_0} \sim a^{-r}, r>0$, as $a\ll1$. Such bubbles are not known to exist in nature but they might be designed, see \cite{Z-F-2018}. These properties mean that \emph{the relative speed of propagation is small}. But the contrast of the transmission coefficient is moderate.
\end{enumerate}
\end{definition}

 \bigskip

\section{Characterization of the key resonances}\label{section3}

From the above modeling of the nano-particles as well the micro-bubbles, we observe that they enjoy one of the following properties:
\begin{enumerate}
 \item the speed of propagation is small and the contrast of the transmission coefficient is moderate.
In this case, even though the particle is small, the wave might spend some time inside it, i.e. we might have local body mode (or vibration) if the speed is small under some critical scales. However, there is no surface mode
as the transmission coefficient is moderate.

\item the speed of propagation is moderate (not small) and the transmission coefficient is large (or singular as the Drude model). In this case, it is the other way around. We might have surface modes under critical scales of the transmission coefficients but it is unlikely to have body modes as the speed is not small enough. 
\end{enumerate}

 These differences give rise to different types of resonances for our two types of materials (micro-bubbles and nano-particles). We classify them as follows

\bigskip

\begin{enumerate}

 \item In the case of Micro-bubbles, we have
 
 \begin{enumerate}
  \item the Minnaert resonance which corresponds to a surface-mode for the Low Density / Low Bulk bubbles.  
  
  \item A sequence of resonances which correspond to body-modes for the Moderate Density / Low Bulk bubbles. 
  \end{enumerate}
\bigskip	
	
\item In the case of nano-particles, 
	
	\begin{enumerate}
	\item the plasmonic sequence of resonances which corresponds to surface-modes for plasmonic nano-particles. 
	
	\item the Mie (or dielectric) sequence of resonances corresponding to body-modes for the dielectric nano-particles.

 \end{enumerate}
\end{enumerate}
Such resonances manifest themselves for special values of $r$. Indeed for $r<2$, they are very large and for $r>2$ they are very small in terms of the relative diameter $a$, $a\ll 1$. However, for $r=2$, they are moderate and their dominant parts are independent of $a$. In the next section, we consider the case $r=2$ and show formally how these resonances indeed appear. For simplicity of the exposition, we handle the cases where the background is homogeneous, i.e. the material coefficients are all constant outside the injected particles. However, the same results occur even for heterogeneous backgrounds.
\subsection{ Formal characterization of the resonances. The acoustic model}

Let $D = z + a\; B$ be a bounded, $C^1$-smooth \footnote{In most of the computations, the Lipschitz regularity is enough. The only place where we need more regularity is in the characterization of the spectrum of the double layer operator.} and connected subset containing the origin of $\mathbb{R}^{3}$, with a 'radius' $a \ll1$.
Let $u = u^s + u^i$ be the solution of the acoustic scattering problem, see \cite{Papa6, Papa7, AMMARI20192104, Habib-Minnaert},
\begin{equation*}\label{eq:acoustic_scattering}
\begin{cases}
\div \frac{1}{\rho} \nabla u + \omega^2 \frac{1}{k} u =  0 \quad \text{in} \; \mathbb{R}^{3},\\

u^s:= u-u^i \text{ satisfies the following Sommerfeld Radiation Conditions (S.R.C.)}\\

\frac{\partial u^{s}}{\partial |x|}-i\; \omega \sqrt{\frac{\rho}{k}}\; u^{s}=o\left(\frac{1}{|x|}\right), |x|\rightarrow\infty.\end{cases}
\end{equation*}

where \begin{equation*}
\label{rho}
\rho:= \begin{cases}
\rho_1 \quad \text{inside } D,\\
\rho_0 \quad \text{outside } D \end{cases}
\text{and} \qquad
k:= \begin{cases}
k_1 \quad \text{inside } D,\\
k_0 \quad \text{outside } D. \end{cases}\end{equation*}

Here $u^i:=u^i(x, \omega, \theta):=e^{i \omega \sqrt{\frac{\rho_0}{k_0}}\; x\cdot \theta}$ is an incident plane wave propagating in the direction $\theta$.

 From the Lippmann-Schwinger representation of the total acoustic field $u$, we have
\begin{equation}
\label{eq:1}
 u(x) - \alpha \, \underset{x}{\div}  \int_D G_\omega (x-y) \nabla u(y) dy -\beta \omega^2 \int_D G_\omega (x-y)  u(y) dy = u^i(x),
\end{equation}
where $\alpha := \frac{1}{\rho_1} - \frac{1}{\rho_0}$ and $\beta := \frac{1}{k_1} - \frac{1}{k_0}$ represent the contrasts  between the inner and the outer acoustic coefficients. Here, $G_\omega$ is the Green's function of the background medium $(\rho_0, k_0)$ satisfying the outgoing Sommerfeld radiation conditions. This is an integro-differential equation. To transform it to a solely integral equation, we proceed by integration by parts then \eqref{eq:1} becomes 
\begin{equation*}
\label{eq:ReworkLippmannSchwinger}
  u(x) -  \gamma \omega^2  \int_D G_\omega(x-y)u(y) dy + \alpha \int_{\partial D} G_\omega (x-y) \frac{\partial u}{\partial \nu} (y) dy = u^i(x) ,
\end{equation*}
where $\gamma := \beta - \alpha \rho_1/k_1$, for $x \in D$. In addition, taking the normal derivative and trace, with the usual traces of the double layer potential, we obtain:

\begin{equation*}
\label{eq:ReworkLippmannSchwinger2}
(1 + \frac{\alpha}{2}) \frac{\partial u}{\partial \nu} - \gamma \omega^2   \partial_{\nu-} \int_D G_\omega(x-y)u(y) dy + \alpha( K_D^{\omega})^* [\frac{\partial u}{\partial \nu} ] = \frac{\partial u^i}{\partial \nu}.
\end{equation*}

Hence for $x\in \mathbb{R}^3\setminus \overline{D}$, the total acoustic field $u(x)$ is characterized by 
$
u|_D $ and $\frac{\partial u}{\partial \nu}|_{\partial D}
$ 
which are solutions of the following close form system of integral equations:
\bigskip

\begin{equation*}
\label{eq:ReworkLippmannSchwinger-1-}
  [I -  \gamma \omega^2 N^\omega]u + \alpha \int_{\partial D} G_\omega (x-y) \frac{\partial u}{\partial \nu} (y) dy = u^i(x), \; \mbox{ in } D
\end{equation*}
\begin{equation*}
\label{eq:ReworkLippmannSchwinger2-}
[\frac{1}{\alpha} +\frac{1}{2} + ( K_D^{\omega})^*] [\frac{\partial u}{\partial \nu} ]  - \frac{\gamma}{\alpha} \omega^2   \partial_{\nu-} \int_D G_\omega(x-y)u(y) dy  = \frac{1}{\alpha}\frac{\partial u^i}{\partial \nu}, \; \mbox{ on } \partial D
\end{equation*}

with the Newtonian (a volume-type) operator:
$$
N^{\omega} : L^2(D)\longrightarrow L^2(D),~~~ N^\omega(u)(x):=\int_D G_\omega(x-y)u(y) dy,
$$
with image of $N^\omega$ in $ H^2(D)$, and the Neumann-Poincar\'e (a surface-type) operator\footnote{The notation $p.v$ means the Cauchy principal value.}
$$
(K_D^{\omega})^*: H^{-1/2}(\partial D) \longrightarrow H^{-1/2}(\partial D),~~~ (K_D^{\omega})^*(f)(x):=p.v.\int_{\partial D} \frac{\partial }{\partial \nu_x} G_\omega (x-y) f (y) dy.
$$

\begin{remark}
The following properties are key in estimating the resonances. For $\omega=0$, each of these operators generates a sequence of eigenvalues: $\lambda_m(N^0)\stackrel{m\rightarrow \infty}{\longrightarrow} 0$ and $\sigma_p((K_D^{0})^*)\subset [ -\frac{1}{2}, \frac{1}{2})$. In addition, we have $K^0_D(1)=-\frac{1}{2}$. \emph{These singular values are behind all the used resonances}.
\end{remark}

Indeed, let us recall the system of integral equation and see how it behaves for the scales defining our micro-bubbles:

\begin{equation}\label{eq:ReworkLippmannSchwinger-1-}
  \left[ I -  \gamma ~ \omega^2~ N^\omega \right]u + \alpha \int_{\partial D} G_\omega (x-y) \frac{\partial u}{\partial \nu} (y) dy = u^i(x), \; \mbox{ in } D,
\end{equation}

\begin{equation}\label{eq:ReworkLippmannSchwinger2-}
\left[ \frac{1}{\alpha} +\frac{1}{2} + (K_D^{\omega})^* \right] \left[ \frac{\partial u}{\partial \nu} \right]  - \frac{\gamma}{\alpha} \omega^2   \partial_{\nu-} \int_D G_\omega(x-y)u(y) dy  = \frac{1}{\alpha}\frac{\partial u^i}{\partial \nu}, \; \mbox{ on } \partial D.
\end{equation}

1. For \emph{Low Density / Low Bulk bubbles}, we have $\gamma \sim 1$ and then $\gamma \, \omega^2 \, N^{\omega} \ll1$ as $a\ll1$. Hence, there is no singularity coming from (\ref{eq:ReworkLippmannSchwinger-1-}). But as $\alpha \gg1$, precisely if $\alpha \sim a^{-2}$ as $a\ll1$, then we can excite the eigenvalue $-\frac{1}{2}$ of $K_D^{0}$ and create a singularity in (\ref{eq:ReworkLippmannSchwinger2-}). In this case, we have the \emph{Minnaert resonance} with surface-modes. This resonance was first observed in \cite{Habib-Minnaert} based on indirect integral equation methods. This observation was used for different purposes, see \cite{A-F-O-Y, A-F-L-Y-Z,  H-F-G-L-Z-1, az}.
This result was extended to more general families of micro-bubbles in \cite{AMMARI20192104, ACCS-effective-media}.   
\bigskip

2. For \emph{Moderate Density / Low Bulk bubbles}, we have $\alpha \sim 1$ and then we keep away from the full spectrum of $(K_D^{0})^*$. Hence there is no singularity coming from (\ref{eq:ReworkLippmannSchwinger2-}). But as $\gamma \sim a^{-2}\gg1$, we can excite the eigenvalues of the Newtonian operators $N^0$ and create singularities in (\ref{eq:ReworkLippmannSchwinger-1-}). This gives us a sequence of resonances with volumetric-modes. This was observed in \cite{A.D-F-M-S} and \cite{M-M-S}.
\bigskip

3. Observe that if $\alpha$ is negative (i.e. \emph{negative mass densities}, similar to the Drude model for acoustics for instance) then we could excite the other sequence of eigenvalues of $(K_D^{0})^*$. This gives us another sequence of resonances (i.e. corresponding to the sequence of plasmonics in electromagnetics, see section \ref{Section-Resonances-Maxwell}).
\bigskip

Using Lippmann-Schwinger equations allows to characterize all these resonances and for varying backgrounds (i.e. heterogeneous backgrounds), see \cite{D-G-S-2020}.

\subsection{Formal characterization of the resonances. The electromagnetic model}\label{Section-Resonances-Maxwell}
We deal with \emph{non-magnetic} materials.
The electric field $E = E^s + E^i$ is solution of the electromagnetic scattering problem, see \cite{colton2012inverse, DK},
\begin{equation*}
\label{eq:electromagnetic_scattering}
\begin{cases}
curl\; curl\; E + \omega^2 \, \epsilon \, \mu_0 \, E =  0 \quad \text{in } \mathbb{R}^{3},\\

E^s:=E-E^i \text{ satisfies the following Silver-Mueller Radiation Condition (S-M.R.C.)}\\

\underset{\vert x \vert\rightarrow\infty}{\lim}\left(curl\; E^s(x)\times x-i \; \omega \sqrt{\epsilon\; \mu_0}\; \vert x \vert \;  E^s(x)\right)=0
\end{cases}
\end{equation*}
where
\begin{equation*}
\epsilon:= \begin{cases}
\epsilon_1 \quad \text{inside } D,\\

\epsilon_0 \quad \text{outside } D. \end{cases}\end{equation*}

Here, $E^i$ is a polarized incident electric field propagating in the background medium. In particular, it is divergence free, i.e. $\nabla \cdot E^i=0$ in the whole space.
\bigskip

The corresponding Lippmann-Schwinger equation is:
\begin{equation}
\label{eq:1-2}
 E(x) + \eta \; \underset{x}{\nabla} \; \int_D \underset{y}{\nabla} G_\omega (x-y)\cdot E(y) dy -\omega^2 \eta \int_D G_\omega (x-y)  E(y) dy = E^i(x),
\end{equation}
where $\eta := \epsilon_1 - \epsilon_0$ is the contrast of the inner and outer electric permittivities. Here, $G_\omega$ is the Green's function of the Helmholtz equation $(\Delta +\omega^2 \epsilon_0\; \mu_0)=-\delta,$ satisfying the Sommerfeld outgoing radiation condition. By integration by parts, \eqref{eq:1-2} becomes 
\begin{equation*}
\label{eq:ReworkLippmannSchwinger-2}
  E(x) -  \eta \omega^2  \int_D G_\omega(x-y)E(y) dy + \eta \nabla \int_{\partial D} G_\omega (x-y) E \cdot \nu (y) dy = E^i(x).
\end{equation*}
 In addition, taking the normal limit from inside $D$, we obtain:
\begin{equation*}
\label{eq:ReworkLippmannSchwinger2-2}
(1 + \frac{\eta }{2}) E\cdot \nu - \eta \, \omega^2  \, \nu \cdot \int_D G_\omega(x-y)E(y) dy + \eta( K_D^{\omega})^* [E\cdot \nu ] = E^i \cdot  \nu, \; \mbox{ on } \partial D.
\end{equation*}


Hence for $x\in \mathbb{R}^3\setminus \overline{D}$, $E(x)$ is characterized by $E|_D$ and $E \cdot \nu|_{\partial D}$ which are solutions of the close system of integral equation:
\begin{equation}\label{eq:ReworkLippmannSchwinger-1-2}
  \left[ I - \eta \omega^2 N^{\omega} \right]E + \eta \, \nabla \int_{\partial D} G_\omega (x-y) E \cdot \nu (y) dy = E^i(x), \; \mbox{ in } D
\end{equation}
and 
\begin{equation}\label{eq:ReworkLippmannSchwinger2-2}
\left[ \frac{1}{\eta} +\frac{1}{2} + (K_D^{\omega})^* \right] [E \cdot \nu]  - \omega^2   \nu \cdot \int_D G_\omega(x-y)E(y) dy  = \frac{1}{\eta} \; E^i \cdot \nu,  \; \mbox{ on } \partial D.
\end{equation}

1. For \emph{dielectric nano-particles}, we have $\eta \gg1 $. Hence $\frac{1}{\eta} \ll1$, however we keep away from the full spectrum of $(K_D^{0})^*$ as the dominating term of the sources is $\frac{1}{\eta} \; E^i \cdot \nu$ which is average-zero as $\nabla \cdot E^i=0$. Hence no singularity comes form (\ref{eq:ReworkLippmannSchwinger2-2}). But if in addition $\eta \sim a^{-2}\gg1$, then we can excite the eigenvalues of the Newtonian operators $N^{0}$ and create singularities in (\ref{eq:ReworkLippmannSchwinger-1-2}). Observe here that the Newtonian operator is vectorial in contrast to the acoustic case where it is a scalar operator. Therefore its spectral decomposition is more involved. Nevertheless, projecting it on the subspace of divergence free fields in $D$ with zero normal component on $\partial D$, we obtain the sequence of dielectric (or Mie) resonances. 
\bigskip

2. Observe that if the real part of $\eta$ is negative (i.e. negative contrast permittivity) as in the Drude model, where $\eta=\epsilon-\epsilon_0=-\frac{\omega_p^{2}}{\omega(\omega +i\; \gamma)}$, then we can excite the sequence of eigenvalues of $(K_D^{0})^*$, which lie in $[ -\frac{1}{2}, \; \frac{1}{2})$, by appropriately choosing the damping parameter $\gamma$ and the incident frequency $\omega$. This gives us the sequence of electric \emph{plasmonics} which create singularities in (\ref{eq:ReworkLippmannSchwinger2-2}). With such material contrast $\eta$, there is no singularity that can be created in (\ref{eq:ReworkLippmannSchwinger-1-2}),
since $N^{\omega}$ scales as $a^2,\; a\ll 1$.

\bigskip

3. It is legitimate to ask if any resonance, corresponding to the Minnaert one in Acoustics, exists? To answer to this question, we consider \emph{magnetic materials}, i.e. $\mu_1 \neq \mu_0$.
In the TM-approximation, the magnetic component $u:=H_3$, satisfies the acoustic scattering problem
\begin{equation*}
\label{eq:acoustic_scattering}
\begin{cases}
\div \frac{1}{\epsilon} \nabla u + \omega^2 \mu u =  0 \quad \text{in } \mathbb{R}^{2},\\
u^s \text{ satisfies the S.R.C}
\end{cases}
\end{equation*}
where $\epsilon = \epsilon_1$ inside $D$, $\epsilon = \epsilon_0=1$ outside $D$, $ \mu = \mu_1$ inside $D$, $\mu = \mu_0=1$ outside $D$. As in the Acoustic case, for $x\in \mathbb{R}^3\setminus \overline{D}$, $u(x)$ is characterized by~~ $u|_D$~~ and~~ $\frac{\partial u}{\partial \nu}|_{\partial D}$~~ which satisfy the system:

\begin{equation*}\label{eq:ReworkLippmannSchwinger-1-}
\begin{cases}
  \left[I -  \tilde{\gamma} \, \omega^2 \,  N^{\omega} \right]u + \tilde{\alpha} \int_{\partial D} G_\omega (x-y) \frac{\partial u}{\partial \nu} (y) dy = u^i(x), \; \mbox{ in } D\\
\, \\
\left[ \frac{1}{\tilde{\alpha}} +\frac{1}{2} + ( K_D^{\omega})^* \right] [\frac{\partial u}{\partial \nu} ]  - \frac{\gamma}{\tilde{\alpha}} \omega^2   \partial_{\nu-} \int_D G_\omega(x-y)u(y) dy  = \frac{1}{\tilde{\alpha}}\frac{\partial u^i}{\partial \nu}, \; \mbox{ on } \partial D
\end{cases}
\end{equation*}

 where, now, $\tilde{\gamma}:= \frac{\epsilon_1 \mu_1}{\epsilon_0}-\mu_0  \mbox{~~ and~~ } \tilde{\alpha}:=\epsilon^{-1}_1-\epsilon^{-1}_0.$

\begin{enumerate}
\item Hence $\tilde{\gamma} \sim 1$ if the index of refraction is moderate, i.e. $\epsilon_1 \mu_1 \sim 1$. Then $\gamma \, \omega^2 \, N^{\omega} \ll1$ as $a\ll1$. 

\item In addition $\tilde{\alpha} \gg1$ if $\epsilon_1 \ll1$, as $a\ll1$. Then we can excite the eigenvalue $-\frac{1}{2}$ of $K_D^{0}$. 
\end{enumerate}
\bigskip

 Hence, \emph{a Minnaert-like resonance can be excited by $\epsilon$-near-zero and $\mu$-near-infinity nano-particles.} The last issue is whether such materials exist in nature or possible to be designed. Material with $\epsilon$-near-zero are possible with the Drude model where $\epsilon:=\epsilon_0-\frac{\omega^{2}_p}{\omega(\omega+i\; \gamma)}$ choosing $\gamma, \gamma \ll 1$, and $\omega$ so that $\epsilon$ is near zero.

\subsection{ Summary on the existence of the resonances}
We summarize here the possibilities we have in creating resonances.

\begin{enumerate}
\item Acoustic bubbles:
\bigskip

\begin{itemize}
\item For low density / low bulk bubbles, we have the Minnaert resonance with surface-modes.

\item For moderate density / low bulk bubbles, we have a sequence of resonances with volumetric-modes.

\item For negative contrasts of mass densities, we have a sequence of resonances (corresponding to the plasmonics in electromagnetism) with surface-modes.

\end{itemize}
\bigskip

\item Electromagnetic nano-particles:
\bigskip

\begin{itemize}

\item For dielectric nano-particles, we have the sequence of Mie (or dielectric) resonances with volumetric-modes.

\item For negative contrast of the permittivity, as Drude's model, we have the sequence of plasmonic resonances with surface-modes.

\item For $\epsilon$-near-zero and $\mu$-near-infinity nano-particles, we have one more resonance (corresponding to the Minnaert one in acoustic bubbles) with surface-modes.

\end{itemize}

\end{enumerate}

\section{Acoustic imaging using resonating micro-bubbles}\label{section4}

\subsection{Expansion of the fields}

Let $D:=z+a\; B$ be the bubble of center $z$ injected in the body to image $\Omega$ which is a bounded and smooth domain in $\mathbb{R}^3$.
\bigskip

Let $v:=v^s +v^{i}$ be the total field generated by the background $(\rho_0, k_0)$ \emph{without the bubble}. Here the coefficients $(\rho_0, k_0)$ are variable and $W^1$-smooth inside $\Omega$ and $(\rho_0, k_0):=(\rho_{0, \infty}, k_{0, \infty})$ outside $\Omega$ where $\rho_{0, \infty}$ and $k_{0, \infty}$ are positive constants.  
\bigskip

We set $u:=u^s+u^i$ to be the total field generated by the background $(\rho, k)$ in \emph{the presence of one bubble}. This means that $(\rho, k)=(\rho_1, k_1)$ inside the bubble $D$ and $(\rho, k)=(\rho_0, k_0)$ outside of it. Here $v^i=u^i:=e^{i\kappa_0 \theta \cdot x}$ is the incident plane wave where $\kappa_0:=\omega\sqrt{\frac{\rho_{0, \infty}}{k_{0, \infty}}}$
and $\theta$ is the direction of incidence. Hence, the acoustic model in the presence of one bubble injected in the heterogeneous background $\Omega$ reads as follows, see \cite{Papa6, Papa7, AMMARI20192104, Habib-Minnaert}, 
\begin{equation*}
\label{eq:acoustic_scattering}
\begin{cases}
\div \frac{1}{\rho} \nabla u + \omega^2 \frac{1}{k} u =  0 \quad \text{in } \mathbb{R}^{3},\\
u^s:= u-u^i \text{ satisfies (S.R.C.).}\end{cases}
\end{equation*}
Due to the Sommerfeld radiation condition, we have the following behavior of the scattered field $u^{s}(x,\theta)$ far away from the target region $\Omega$:
\begin{equation*}
u^{s}(x,\theta)= \frac{e^{i \kappa_{0} \vert x \vert}}{\vert x \vert} u^{\infty}(\hat{x}, \theta)  + \mathcal{O}\left(\vert x \vert^{-2}\right), \;\; |x| \to + \infty,
\end{equation*}
where $\hat{x}:=x / |x| $ and $u^{\infty}(\hat{x}, \theta) $ denotes the far-field pattern corresponding to the unit vectors $\hat{x},\theta $, i.e. the incident and propagation directions respectively.

Here we take the scales $
k_1:=\bar{k}_1\; a^2 ~\mbox{  and } ~\rho_1:=\bar{\rho}_1\; a^2,$ where 
$\bar{k}_1$ and $\bar{\rho}_1$ are positive constants independent of $a$. With these scales, we have existence of \emph{the Minnaert resonance}. Indeed, we have the expansion, see \cite{D-G-S-2020}:
\begin{equation}\label{scatered-field-final-form-Minnaert}
 u^\infty(\hat{x}, \theta, \omega, z) =v^\infty(\hat{x}, \theta, \omega)-\frac{1}{\overline{k}_{1}}\; \frac{\omega^2_M}{\omega^2-\omega^2_M}\vert B \vert\; a \; 
v(z, -\hat{x}, \omega)\; v(z, \theta, \omega) + O \left( \frac{a^{2}}{\left( \omega^{2} - \omega^{2}_{M}  \right)^{2}} \right)
\end{equation}
where 
\begin{equation}\label{Minnaert-resonance}
 \omega_M=\omega_M(z):=\sqrt{\frac{8 \, \pi \, \overline{k}_{1}}{\rho_0(z) \, A_{\partial B}}}~~~ (\emph{~The~Minnaert~  resonance!})
\end{equation}
with $A_{\partial B}:= \frac{1}{\left\vert \partial B \right\vert} \int_{\partial B}  \int_{\partial B} \frac{(x-y) \cdot \nu(x)}{4 \pi |x-y| } dx dy.$
\bigskip

The far-field $u^\infty(\hat{x}, \theta, \omega, z)$ depends on the frequency of incidence $\omega$ and also the injected bubble at the point $z$, that is why we added the parameter $z$. The far-field $v^\infty(\hat{x}, \theta, \omega)$ corresponds to scattering by the background but without the bubble, hence the parameter $z$ is not added.

\subsection{ Solution of the inverse problem using one injected bubble}
The acoustic imaging problem we are interested in is to reconstruct the background coefficients $k_0$ and $\rho_0$ inside the imaging target $\Omega$. For this, we need the following data.
\begin{enumerate}
\item The back-scattered far-field $v^\infty(-\theta, \theta, \omega)$ in one single direction $\theta$ measured before injecting any bubble.

\item The back-scattered far-field $u^\infty(-\theta, \theta, \omega, z)$ in one single direction $\theta$ measured for each injected bubble, located in a point $z \in \Omega$.
\end{enumerate}

We use these data for a band of frequencies $$[\omega^{min}_M,~~ \omega^{max}_M]$$ where 
$$
\omega^{min}_M < \sqrt{ \frac{4 \, \pi \,\overline{k}_{1}}{\Big(\underset{\Omega}{\max}\, \rho_0(z) \Big)~\;A_{\partial B}}}\; \mbox{ and }\; \omega^{max}_M> \sqrt{\frac{4 \, \pi \,  \overline{k}_{1}}{\Big(\underset{\Omega}{\min} \, \rho_0(z)\Big)\;~A_{\partial B}}}.
$$

The imaging procedure goes as follows. We set 
\begin{equation}\label{Imaging-functional}
I(\omega, z):=u^\infty(-\theta, \theta, \omega, z)-v^\infty(-\theta, \theta, \omega)
\end{equation}
 as the imaging functional, remembering that the incident angle $\theta$ is fixed. We have the following properties from (\ref{scatered-field-final-form-Minnaert})
\begin{equation}\label{imaging-functional}
I(\omega, z) \sim -\frac{\omega^2_M}{ \overline{k}_1 (\omega^2-\omega^2_M(z))}\vert B \vert\; a \;\; [v(z, \theta, \omega)]^2.
\end{equation}
We divide this procedure into two steps:
\begin{enumerate} 
\item Step 1. From this expansion, we recover $\omega^2_M(z)$ as the frequency for which the imaging function $\omega \rightarrow I(\omega, z)$ gets its largest value. From the estimation of this resonance $\omega^2_M(z)$, we reconstruct the mass density at the center of the injected bubble $z$, based on (\ref{scatered-field-final-form-Minnaert}), as follows:
\begin{equation*}\label{reconstruction-rho}
\rho_0(z)= \frac{4 \pi\, \overline{k}_{1}}{ \omega^2_M(z) \; A_{\partial B}}. 
\end{equation*}
Scanning the domain $\Omega$ by such bubbles, we can estimate the mass density there. 
\bigskip
\item Step 2. To estimate now the bulk modulus, we go back to (\ref{imaging-functional}) and derive the values of the total field $[v(z, \theta, \omega)]^2$. This field corresponds to the model without the bubble. 
Hence, we have at hand $v(z, \theta, \omega)$ for $z \in \Omega$ up to a sign (i.e. we know the modulus and the phase up to a multiple of $\pi$). Use the equation $\nabla\cdot \rho^{-1}_0 \nabla v +\omega^2 k^{-1}_0 v=0$ to recover the values of $k_0$ in the regions where $v$ does not change sign. This can be done by numerical differentiation for instance. Other ways are of course possible to achieve this second step. In addition, we have at hand multiple frequency internal data.

\end{enumerate}
\bigskip

The procedure described above uses the Minnaert resonance. The key point to recover the mass density is the explicit dependance of this resonance on the value of the mass density, of the background, on it's 'center', see (\ref{Minnaert-resonance}).

\begin{remark} We have shown that using multifrequency back-scattering data at one single frequency, we can reconstruct the mass density with a simple and explicit formula. In addition, we can estimate the internal values of total field from which we derive the values of the bulk modulus via numerical differentiation. Disadvantages of this procedure are as follows:  

\begin{itemize}

\item Possible zeros of the total fields $v(z, \theta, \omega_M(z))$. Indeed, we cannot recover the bulk modulus $k_0$ in the regions where this total field vanishes. But this is natural and cannot be avoided without extra information. One way to handle this issue is to use multiple directions of incidence $\theta$.
\bigskip

\item Numerical differentiation.  The numerical differentiation is an unstable step. In addition, we need to differentiate twice reconstructed, and hence noisy, quantities given by $\rho_0$ and $v$. This makes it a difficult issue to handle in practice. One way to remedy to this is to use two injected bubbles which are close to each other. Doing so, we can recover not only the total field $v$ but also the Green's function $G_{\omega_M}$, of the background medium, on the 'centers' of the two bubbles. From the singularities of $G_{\omega_M}$, we recover the bulk $k_0$. The idea of the proof is based  on the \emph{Foldy-Lax paradigm} on whether one can extract the multiple scattering field generated by closely spaced small particles from the measured far-field. This paradigm is justified for nearly resonating frequencies. This argument was tested in a similar situation, see \cite{A-A-C-K-S}. A complete study of this paradigm for our settings is done in \cite{G-S-2021}. 
\end{itemize}

\end{remark}

\subsection{Summary on the Acoustic imaging using resonating bubbles}
Here, we summarize the way how we propose to solve the acoustic imaging problem. 
\begin{enumerate}

\item  Injecting single bubbles and using the generated back-scattered field in one incident direction, sent at multiple frequencies, 
we can reconstruct 
\bigskip

\begin{enumerate}
\item the density $\rho_0$ via direct and stable formulas,
\bigskip

\item the bulk $k_0$ with numerical differentiation.
\end{enumerate}
\bigskip

\item Injecting double and close bubbles (i.e. dimers), we can avoid the numerical differentiation.

\end{enumerate}

\section{Photo-acoustic imaging using resonating nano-particles}\label{section5}
\subsection{The mathematical model}
The photo-acoustic experiment, in the general setting, applies to targets that are electrically conducting, in other words the imaginary part of the 'permittivity' is quite pronounce, and it goes as follows. Exciting the target, with laser, or by sending an incident electric field, will create heat in it surrounding. This heat, in its turn, creates fluctuations, i.e. a pressure field, that propagates along the body to image. This pressure can be collected in an accessible part of the boundary of the target. The photo-acoustic imaging is to trace back the pressure and reconstruct the permittivity that created it.       

To describe the mathematical model behind this experiment, let us set $E$, $\bm{T}$ and $p$ to be respectively the electric field, the heat temperature and the acoustic pressure. Then, as described above, the photo-acoustic experiment is based on the following model coupling these three quantities:  

\begin{equation*}\label{Photoacoustic-general-model}
\left \{
\begin{array}{llrr}

curl\; curl\; E -\omega^2\; \varepsilon\; \mu_0\; E=0,~~~ E:=E^s+E^i, \mbox{ in } \mathbb{R}^{3},\\
\\
\rho_0 c_p\dfrac{\partial \bm{T}}{\partial t}-\nabla \cdot \kappa \nabla \bm{T} =\omega\; \Im(\varepsilon)\;\vert E \vert^2\; \delta_{0}(t),\; \mbox{ in } \mathbb{R}^{3}\times \mathbb{R}_+,\\
\\
\dfrac{1}{c^2}\dfrac{\partial^2 p}{\partial t^2}-\Delta p= \rho_0\; \beta_0\; \dfrac{\partial^2 \bm{T}}{\partial t^2}, \mbox{ in } \mathbb{R}^{3}\times \mathbb{R}_+,
\end{array} \right.
\end{equation*}
 where $\rho_0$ is the mass density, $c_p$ the heat capacity, $\kappa$ is the heat conductivity, $c$ is the wave speed and $\beta_0$ the thermal expansion coefficient. 
To the last two equations, we supplement the homogeneous initial conditions: 
\begin{equation*}
 \bm{T} =p=\frac{\partial p}{\partial t}=0, \mbox{ at } t=0
\end{equation*}
and the Silver-Mueller radiation condition to $E^s$.
More details on the actual derivation of this model can be found in \cite{P-P-B:2015, Triki-Vauthrin:2017} and more references therein. 
\bigskip

In our settings, the source of the heat is given by the injected electromagnetic nano-particles. Precisely, we propose to inject dielectric nano-particles. As described in section \ref{section2} and section \ref{section3}, these nano-particles enjoy the following features. They are highly localized as they are nano-scaled and they have high contrast permittivity. Under critical scales, that will be described later, these two features allow us to estimate the dominant part of the measure pressure and then extract from it the unknown permittivity, and eventually the conductivity as well. 
\bigskip

To give more details on how such procedure works, we restrict ourselves to
\bigskip

\begin{enumerate}
 \item  the 2D-TM model for the electromagnetic propagation.
\bigskip

\item the heat conductivity $\kappa$ is small and can be neglected.  
\end{enumerate}
\bigskip

Under such conditions, the $2D$ model of the photo-acoustic reduces to one single equation, see \cite{Triki-Vauthrin:2017}. 

\begin{equation}\label{pressurwaveequa}
\left\{
\begin{array}{rll}
    \partial^{2}_{t} p(x,t) - c_{s}^{2}(x) \Delta_{x} p(x,t) &=& 0 \qquad in \quad \mathbb{R}^{2} \times \mathbb{R}^{+}.\\
    p(x,0) &=& \frac{\omega \, \beta_{0}}{c_{p}} \Im(\varepsilon)(x) \, \vert E_3 \vert^{2}(x), \qquad in \quad \mathbb{R}^{2} \\ 
    \partial_{t}p(x,0) &=& 0 \qquad in \quad \mathbb{R}^{2} 
    \end{array}
\right.
\end{equation}
here $c_{s}$ is the velocity of sound in the medium that is smooth and $c_s-1$ is supported in a smooth and compact set $\Omega$. The constants $\beta_0$ and $c_p$ are known and $\omega$ is an incident frequency. The source $u:=E_3$ is solution of the scattering problem
\begin{equation}
\label{eq:electromagnetic_scattering}
\left\{
\begin{array}{rll}
\Delta u + \omega^2 \mu_0 \epsilon\;  u =  0 \quad \text{in } \mathbb{R}^{2},\\
\\
u(x):=u^s(x)+e^{i\; \omega \sqrt{\mu_0 \epsilon_0}\;x }\\
\\
u^s \text{ satisfies the S.R.C.}
\end{array}
\right.
\end{equation}
where $\epsilon = \epsilon_1$ inside $D$, $\epsilon = \epsilon_0$ outside $D$ and $\epsilon_0=1$ outside $\Omega$
 ($D \subset \Omega$ being the injected nano-particle). The permittivity $\epsilon_0$ is variable and it  is supposed to be Lipschitz smooth inside $\Omega$.

\subsection{ Inversion of the photo-acoustics using nano-particles}
The goal of the photo-acoustic imaging using nano-particles is to recover $\epsilon$ in $\Omega$ from the measure of the pressure $p(x, t),\; x \in \partial \Omega$ and $t \in (0, T)$ for large enough $T$. The decoupling of the original photo-acoustic mathematical model (\ref{Photoacoustic-general-model}) into (\ref{pressurwaveequa})-(\ref{eq:electromagnetic_scattering}) suggests that we split the inversion into the following two steps.

\begin{enumerate}
\item Acoustic Inversion: Recover the source term $\Im(\varepsilon)(x) \, \vert u \vert^{2}(x)$, $x \in \Omega$, from the measure of the pressure $p(x, t),\; x \in \partial \Omega$ and $t \in (0, T)$.
\bigskip

\item Electromagnetic Inversion: Recover the permittivity $\epsilon(x),\; x\in \Omega$ from  $\Im(\varepsilon)(x) \, \vert u \vert^{2}(x)$, $x \in \Omega$. 
\end{enumerate}
\bigskip

The pressure is collected on the boundary of $\Omega$ in the following situations:
\bigskip

\begin{itemize}

\item Before injecting any particle.~~ There is a considerable amount of works in the literature based on such data. Without being exhaustive, we cite the following references \cite{Habib-book, B-E-K-S:2018, B-G-S:2016, B:2014, B-B-M-T, B-U:2010, C-A-B:2007, FHR, Kirsch-Scherzer, K-K:2010, KuchmentKunyansky, N-S:2014, Natterer, S:2010, S-U:2009} devoted to such inversions.
\bigskip

\item After injecting a single particle. To our best knowledge, there is only the work \cite{Triki-Vauthrin:2017} where plasmonic nano-particles  are used and an optimization method was proposed to invert the electric energy fields. There, the $2D$-model is stated and the magnetic field was used.  
\bigskip

\item After injecting a double and close particles (dimers).
\bigskip

\item After injecting a cluster of particles. 

\end{itemize}

In the sequel, we show how we can use the data given by the second, third, or forth possibilities to solve the photo-acoustic imaging problem using dielectric nano-particles.

\subsection{ Some Known Acoustic Inversions \cite{K-K:2010, KuchmentKunyansky, Natterer}.}
Here, we describe two different methods proposed to do the inversion of the acoustic field and recover the initial data. The first one is related to the Radon transform and the second one is based on spectral theory.

\begin{enumerate}

\item If the speed of propagation $c_s$ is constant and $\Omega$ is a disc of radius $R$, then
\begin{equation*}
\Im(\varepsilon)(x) \, \vert u \vert^{2}(x) = \frac{1}{2 \pi R}  \int_{\partial \Omega} \int_{0}^{2R} (\partial_{r} \, r \, \partial_{r} M(\Im(\varepsilon) \, \vert u \vert^{2}))(p,r) \, \log(\vert r^{2} - \vert x-p \vert^{2} \vert) \, dr \, d\sigma(p) 
\end{equation*} 
where
\begin{equation*}\label{Abelequa}
M(\Im(\varepsilon) \, \vert u \vert^{2})(x,r) = \frac{2 \omega \beta_{0}}{c_{p} \pi} \int_{0}^{c_{s}r} \frac{p(x,t)}{\sqrt{r^{2}-t^{2}}} dt.
\end{equation*}
See \cite{Natterer} for more details and \cite{FHR}, and the references therein, for further development on the inversion of the Radon transform.
\item Otherwise, but under certain conditions as the non-trapping one, we have the spectral representation
\begin{equation*}
\Im(\varepsilon)(x) \, \vert u \vert^{2}(x) = \frac{c_{p}}{\omega \, \beta_{0}} \, \sum_{k} (\Im(\varepsilon)(x) \, \vert u \vert^{2})_{k} \psi_{k}(x)
\end{equation*}
where
\begin{equation*}
(\Im(\varepsilon)(x) \, \vert u \vert^{2})_{k} = \lambda^{-2}_{k} g_{k}(0) - \lambda^{-3}_{k} \int_{0}^{\infty} \sin(\lambda_{k}t) \, g^{\prime \prime}_{k}(t) dt
\end{equation*}
with 
\begin{equation*}
g_{k}(t) = \int_{S} p(x,t) \frac{\partial \overline{\psi_{k}}}{\partial \nu}(x) dx
\end{equation*}
and $(\lambda_k, \psi_k)$ is the eigen-system of $-c^{-2}_{s}(x) \Delta$ with zero Dirichlet boundary condition on $\partial \Omega$. See \cite{K-K:2010, KuchmentKunyansky} for more details.
\end{enumerate} 

\subsection{ An Approximative Acoustic Inversion  \cite{AhceneMourad}.}

Contrary to the previous described results, here we need no condition on the geometry of the shape of $\Omega$ nor a trapping conditions. We assume the permittivity $\epsilon_0(\cdot)$, of the medium, to be $W^{1, \infty}-$smooth in $\Omega$ and the permeability $\mu_0$ to be constant and positive. We propose approximative inversions of the acoustic pressure fields to retrieve the initial data under some conditions on localization and large contrast of the permittivity of the injected dielectric nano-particles. Precisely, we assume that the injected nano-particles enjoy the properties: $$\Re \left( \epsilon_p \right) \sim a^{-2} \vert \log(a) \vert^{-1}
~~~\mbox{and}~~~ \Im \left( \epsilon_p \right) \sim a^{-2} \vert \log(a) \vert^{-1-h-s},\;~~~ s > 0.$$

The frequency of the incidence $\omega$ is chosen close to the dielectric resonance $\omega_{n_0}$: 
$$\omega^2_{n_0} := \left( \mu_0 \, \epsilon_p \, \lambda_{n_0} \right)^{-1},$$ as follows
\begin{equation}\label{resoance-n-0-2}
\omega^2:=\omega_{\pm}^2:=\Re(\omega^2_{n_0})(1\pm \vert \log(a)\vert^{-h}),\;~~ 0<h<1
\end{equation}
where $\lambda_{n_0}$ is an eigenvalue of the Newtonian operator acting as: 
$$
A_0u(x):=\int_D -\frac{1}{2\pi} \ln(\vert x -y\vert)u(y)dy.
$$

Let $x \in \partial \Omega$ and $t \geq  diam(\Omega)$. Under the condition $0 < s < 1-h $, we have the following expansions of the pressure:
\bigskip

\begin{enumerate}

\item {\it{Injecting one nano-particle}}. In this case, we have the expansion

\begin{equation*}\label{pressure-to-v_0}
(p^{+} + p^{-} - 2p_{0})(t,x)  =  \frac{-t \, \omega \, \beta_{0}}{c_{p} \, (t^{2}-\vert x-z \vert^{2})^{3/2}}\;   2\; \Im(\epsilon_p)  \int_{D} \vert u_1(x)\vert ^2dx
 + \mathcal{O}\big(\vert \log(a) \vert^{\max(-1,2h-2)}\big).
\end{equation*}

Here,  $p^+$ and $p^-$ correspond to the frequency $\omega_{+}^2:=\Re(\omega^2_{n_0})(1+ \vert \log(a)\vert^{-h})$, respectively $\omega_{-}^2:=\Re(\omega^2_{n_0})(1- \vert \log(a)\vert^{-h})$, see ({\ref{resoance-n-0-2}}) and
$u_1$ is the total electric field in the presence of one dielectric nano-particle.
\bigskip

\item {\it{Injecting two close dielectric nano-particles.}} We have the following expansion
\begin{equation*}\label{pressure-tilde-expansion}
(p^{+} + p^{-} - 2p_{0})(t,x)  =  \frac{-t \, \omega \, \beta_{0}}{c_{p} \, (t^{2}-\vert x-z \vert^{2})^{3/2}}\;  4 \; \Im(\epsilon_p)  \int_{D} \vert u_2(x)\vert ^2dx
 + \mathcal{O}\big(\vert \log(a) \vert^{\max(-1,2h-2)}\big).
\end{equation*}
where $D$ is any one of the two nano-particles. Here,  $p^+$ and $p^-$ correspond to the frequency $\omega_{+}^2:=\Re(\omega^2_{n_0})(1+ \vert \log(a)\vert^{-h})$, respectively $\omega_{-}^2:=\Re(\omega^2_{n_0})(1- \vert \log(a)\vert^{-h})$, see again ({\ref{resoance-n-0-2}}) and
$u_2$ is the total electric field in the presence of two closely spaced dielectric nano-particles.
\end{enumerate}
\bigskip

Measuring $p^+(x, t),~~ p^-(x, t)$ and~~ $p_0(x, t)$ for two single points $x_1 \neq  x_2~~ \mbox{  in }~~ \partial \Omega~~$  at two single times $t_1\neq t_2$, we can 
\begin{enumerate}

\item localize the center of the injected single nano-particle $z$ and estimate $\int_{D} \vert u_1(x)\vert ^2dx$.

\bigskip

\item estimate the center of the two injected nano-particles $z_1, z_2$ (but we do not distinguish them). In addition,
we can estimate $\int_{D} \vert u_2(x)\vert ^2dx$. Here $D$ is any of the two nano-particles.  
\end{enumerate}
\bigskip

In the next section, we show how we can recover the permittivity $\epsilon_0(z)$ from the previous recovered data, i.e. $\int_{D} \vert u_1(x)\vert ^2dx$ and $\int_{D} \vert u_2(x)\vert ^2dx$. Later, after scanning $\Omega$
with such nano-particles, we can recover $\epsilon_0$ in $\Omega$. We call this step, the electromagnetic inversion.

\subsection{Electromagnetic Inversion}

\subsubsection{Injecting one single particle at once}

We start with the case when we use only single nano-particles as contrast agents. In this case, we have the following approximation
\begin{equation*}\label{one-particle-reconst-permittivity-only}
\int_D\vert u_1\vert^2(x) dx =\frac{\vert u_0(z)\vert^2 (\int_D e_{n_0}(x) dx)^2}{\vert 1-\omega^2 \mu_0 \epsilon_p \lambda_{n_0}\vert^2} + \mathcal{O}\left( a^{2} \, \left\vert \log(a) \, \right\vert^{3h-1} \right).
\end{equation*}

Hence, we can extract the internal phaseless information $\vert u_0(z)\vert$. Recall that $u_0$ is solution of
\begin{equation}\label{phaseless-inverse}
\Delta u_0 +\omega^2\mu_0 \epsilon_0 u_0=0.
\end{equation}

This means that measuring before and after injecting one nano-particle and scanning $\Omega$ with such nano-particles, we transform the photo-acoustic problem to an \emph{inverse problem of reconstructing $\epsilon_0$ from internal phaseless data $\vert u_0\vert$ with $u_0$ solution of (\ref{phaseless-inverse})}. 
\bigskip

Now, we move to the case where we use dimers.

\subsubsection{Injecting double particles at once}
Injecting two closely spaced nano-particles located at $z_1$ and $z_2$.
In this case, we have at hand 
$$
\int_D\vert u_1\vert^2(x) dx ~~~\mbox{ and }~~~ \int_D\vert u_2\vert^2(x) dx.
$$ 

Based on the Foldy-Lax approximation for frequencies near the resonances, see ({\ref{resoance-n-0-2}}), we derive the following expansion

\begin{equation}\label{reconstruction-k-using-contras-permittivity-only}
\log \vert k\vert (z)=  \log(2) - 2\pi \, \gamma^{\star} -
\frac{\frac{\int_D\vert u_1 \vert^2(x) dx}{\int_D\vert u_2 \vert^2(x) dx}-(1-C\Phi_0)^2}{\frac{\int_D\vert u_1 \vert^2(x) dx}{\int_D\vert u_2 \vert^2(x) dx}-2(1-C \Phi_0)}+O(\vert \log(a)\vert^{\max\{h-1, 1-2h\}}),\;~~ a\ll1,
\end{equation}
where $\gamma^{\star}$ is the Euler constant, $\Phi_0:=-\frac{1}{2\pi}\; \ln\vert z_1-z_2\vert$ and 
\begin{equation*}
\textbf{C}:=\int_D[\frac{1}{\omega^2 \mu_0 \Re \epsilon_p}I-A_0]^{-1}(1)(x)dx=\frac{\omega^2 \mu_0 \Re \epsilon_p}{1-\omega^2 \mu_0 \Re\epsilon_p \lambda_{n_0}}\Bigg(\int_D e_{n_0} dx\Bigg)^2 +O(\vert \log(a) \vert^{-1}), \;~~ a\ll1.
\end{equation*}
\bigskip

As we have
\begin{equation*}
\vert k\vert (z)=\omega^2 \vert \epsilon_0\vert \mu_0=\omega^2 \Big(\vert \epsilon_{r}\vert^2 +\frac{\vert \sigma_{\Omega} \vert^2}{\omega^2}\Big)^{1/2} \, \mu_{0}, 
\end{equation*} 
then using two different resonances $\omega_{n_0}$ and $\omega_{n_1}$, \emph{we can reconstruct both the permittivity $\epsilon_r(z)$ and the conductivity $\sigma_{\Omega}(z)$}.
\bigskip

The results described in the two last sections can be found in \cite{AhceneMourad}.
\bigskip

Finally, let us describe the steps needed to follow in doing the electromagnetic imaging using a cluster of injected nano-particles.

\subsubsection{ Imaging using a cluster of contrast agents}

We inject a cluster of contrast agents $(D_m, \epsilon_m, \mu_0), m=1, 2, ..., M$ inside $\Omega$. We need the following assumptions on the distribution of the cluster.
\bigskip

\begin{enumerate}

\item We have both~~~ $\Re \epsilon_m~~ \sim~~ \overline{\epsilon_m}_r a^{-2} $~~~ and~~~ $\Im \epsilon_m~~ \sim~~ \overline{\epsilon_m}_i a^{-2+h}$,~~ with~~ $h \in (0, 1)$. 
\bigskip

\item $1-\frac{\omega^2_{n_0}}{\omega^2}=l_M a^{h}$,~~ with~~ $l_M \neq 0$ and $h \in (0, 1)$.
\bigskip

\item There exists a function $K$ such that
\begin{equation}
\frac{1}{[a^{-1+h}]} \sum^{[a^{-1+h}]}_{j\neq m}\frac{f(z_j)}{\vert z_j-z_m\vert}-\int_{\Omega}\frac{f(z)}{\vert z-z_m\vert}K(z) dz=o(1) \Vert f\Vert_{C^0(\Omega)}, \mbox{uniformly for } z_j \mbox{ and as } a\ll1.
\end{equation}
\end{enumerate}
The first assumption means that we use dielectric nano-particles. Observe that the ration $\frac{\Im \epsilon_m}{\Re \epsilon_m}$ is very small when $a \ll 1$ which mean that the Q-factor (i.e. the quality factor) is very high. This is the most important property that the dielectric nano-particles have. 
The second assumption means that we should use nearly resonating incident frequencies $\omega$. Indeed, this condition is key and it cannot be avoided. The third condition can be quite critical as it means that we have control of the distributed of the nano-particles after injecting them. Nevertheless, as we use a cluster of $M$ particles of the order $M\sim  a^{h-1}$,~~~ $0<h<1$~~ and~~ $a\ll1$, it means that we do not need to inject that many of such nano-particles as soon the used frequency is close to the resonance, i.e. taking $h$ close to 1!
\bigskip

Under these assumptions, we have~~ $
u(x, \theta)-u_K(x, \theta)=o(1), \mbox{ as } a\ll1,
$ where the effective field $u_K$ satisfies the effective problem:  
\begin{equation}\label{Effective-field}
\left(\Delta + \omega^2_{n_0}\epsilon_{K}(x) \mu_0 \right)u^{t}_{K}=0,\ \text{in} \ \mathbb{R}^{3},~~~
u^{t}_{K}=u^{s}_{K}+e^{i \kappa_{0} x \cdot \theta},~~~
\frac{\partial u^{s}_{K}}{\partial \vert x \vert}-i \kappa_{0} u^{s}_{K} =o\left(\frac{1}{\vert x \vert} \right), \ \vert x \vert\rightarrow \infty,
\end{equation}

with ~~~$ \Re \epsilon_K:= \Re \epsilon -K\frac{\vert B \vert}{l_M}\overline{\epsilon_m}_r \chi_{\Omega}. $ ~~~  and~~~ $\Im \epsilon_K=\Im \epsilon + K\frac{\vert B \vert}{l_M}\overline{\epsilon_m}_r \chi_{\Omega}$.
\bigskip

Choosing $l_M>0$, we have
\begin{center}
$\Re \epsilon_K(x)<0,~~~ \mbox{ for }~~~ x \in \Omega ~~~ \mbox{ and }~~~ \epsilon_K(x)-\epsilon \mbox{ can be large }. $
\end{center}

We can use $l_M\ll1$ or $K\gg1$ to enhance these contrasts.
\bigskip

We claim the following steps for reconstructing the permittivity.

\begin{enumerate}
\item From the measured pressure after injecting the cluster, we recover the pressure due to 
the effective medium $\epsilon_K$. The advantage here is that we use a sparse cluster of nano-particles with nearly resonating frequencies however.
\bigskip

\item From this data, we recover $\Im\epsilon_K\; \vert u_K\vert^2$ and hence $\vert u_K\vert^2$ as $\epsilon_K(x)-\epsilon$ is large and known. 
\bigskip

\item This phaseless total internal field corresponds to a locally coercive Helmholtz wave propagator, i.e. (\ref{Effective-field}). By the effective medium theory, we switch the sign of the index of refraction as for metamaterials (in material sciences).  
\bigskip

\item Due to the coercivity, the corresponding least squares functional has a positive second Gateau-derivative. Hence it is a convex functional. Such and observation was already made by I. Knowles, see \cite{K1, K2}.
\bigskip

\item The slope of the least squares functional is sharper as $l_M\ll1$ (or $K\gg1$).

\end{enumerate}

\section{Conclusion}\label{section6}
In this section, we summarize to some extent the discussion we have made in the whole text by emphasizing on the key features and the possible extensions. 

\begin{enumerate}

\item We do believe that imaging with contrast agents is among the promising modalities that are at the cutting edge of modern medical imaging.  
\bigskip

\item We have a clear correspondence between the critical scales of the contrasting materials and the actual resonances.   
\bigskip

\item Using nearly resonating frequencies provides simple and direct links between the measured data and the background coefficients. 
\bigskip

\item We have demonstrated this in two frameworks: Acoustic Bubbles and Electromagnetic Nano-particles in their simplest models however. The original models should be more interesting of course. 
\bigskip

\item Combination of imaging techniques with the effective medium theory might be applied successfully to different modalities as Raman Imaging and Magnetic Resonance Electric Impedance Imaging (MREIT). In particular, regarding MREIT and as stated in the literature, the most famous algorithm, called the harmonic $B_z$ algorithm \cite{LSSW:2007}, and its variants, are based on estimates of the lower bound of Jacobean of the harmonics mappings, see \cite{LSSW:2007}. It is known that this lower bound cannot be achieved in the $3D$ settings. We believe that this shortcoming can be removed as we can retrieve coercivity if we inject resonating nano-particles as contrast agents. However, we need first to revisit the modeling behind and avoid the low frequency approximation that is used so far. In addition, the other imaging modalities as the Magnetic Particle Imaging and Nuclear Imaging (Hadron therapy) would be interesting and challenging as well. 

\end{enumerate}


\end{document}